\documentclass[12pt, letterpaper]{revtex4}
\usepackage{graphicx}
\usepackage{amsmath}
\usepackage{amsfonts}
\usepackage{graphicx}
\usepackage{hyperref}
\usepackage{enumerate}
\usepackage{slashed}

\newcommand{\be}{\begin{equation}}
\newcommand{\ee}{\end{equation}}
\newcommand{\bea}{\begin{eqnarray}}
\newcommand{\eea}{\end{eqnarray}}

\newcommand{\Lag}{\mathcal{L}}
\newcommand{\pd}{\partial}

\newcommand{\matleft}{\left(\begin{array}}
\newcommand{\matright}{\end{array}\right)}

\newcommand{\delete}[1]{}

\begin{document}

\title{Stability of $SU(N_c)$ QCD3 from the $\epsilon$-Expansion}

\author{Hart Goldman$^{1,2}$ and Michael Mulligan$^{2,3}$}
\affiliation{$^{1}$Department of Physics, University of Illinois, Urbana-Champaign, Illinois 61801, USA}
\affiliation{$^{2}$Stanford Institute for Theoretical Physics, Stanford University, Stanford, California 94305, USA}
\affiliation{$^{3}$Department of Physics and Astronomy, University of California, Riverside, California 92511, USA}

\date{\today}

\begin{abstract}
QCD with gauge group $SU(N_c)$ flows to an interacting conformal fixed point in three spacetime dimensions when the number of four-component Dirac fermions $N_f \gg N_c$.
We study the stability of this fixed point via the $\epsilon$-expansion about four dimensions.
We find that when the number of fermions is lowered to $N_f^{\rm crit} \approx {11 \over 2} N_c + (6 + {4 \over N_c}) \epsilon$, a certain four-fermion operator becomes relevant and the theory flows to a new infrared fixed point (massless or massive).
F-theorem or entanglement monotonicity considerations complement our $\epsilon$-expansion calculation.
\end{abstract}

\maketitle

\newpage

\tableofcontents

\section{Introduction and Summary}

Quantum chromodynamics in three spacetime dimensions (QCD3) with gauge group $SU(N_c)$ flows to an interacting conformal fixed point when the number of fermion flavors $N_f \gg N_c$. 
In this paper, we define QCD3 in terms of $N_f$ four-component Dirac fermions in the fundamental representation of $SU(N_c)$.
As the number of fermion flavors is decreased towards $N_c$, the infrared (IR) fate of the theory is less clear.
At what point $N_f^{\rm crit}(N_c) \equiv N_f^{\rm crit}$, if any, does the theory confine or alternatively flow to a new non-trivial IR fixed point?
In this note, we report new progress on this question obtained through an $\epsilon$-expansion about four spacetime dimensions.

The logic of our approach goes as follows.
To leading order in the $\epsilon$-expansion, the QCD3 $\beta_g$-function for the gauge coupling $g$,
\begin{align}
\label{gaugebeta}
\beta_{g} = - {1 \over 2} \epsilon g+ {1 \over 48 \pi^2} (2 N_f - 11 N_c) g^3 + {\cal O}(g^5),
\end{align}
has a non-trivial perturbative zero at $g_\ast^2 = ({24 \pi^2 \over 2 N_f - 11N_c}) \epsilon$ when $N_f > {11 \over 2} N_c$ and $\epsilon \ll 1$ \cite{PeskinSchroeder}.
The $\beta_g$-function in Eq. (\ref{gaugebeta}) indicates that this ``large-$N_f$ fixed point" disappears for $N_f < {11 \over 2} N_c$. 
However, it is possible that the large-$N_f$ fixed point is destabilized earlier by operators that are irrelevant for $N_f > N_f^{\rm crit} > {11 \over 2} N_c$, but become relevant for smaller $N_f$.
If such ``dangerously irrelevant" operators exist (and there is no fine tuning), they necessarily result in the flow to a new IR fixed point that may be either massive or massless.

Under the assumption that the quadratic (mass) perturbations are zero, we find that a linear combination of four-fermion operators (described in Sec. \ref{dangerousirrelevants}) becomes relevant as the number of flavors is lowered towards
\begin{align}
\label{criticalN}
N_f^{\rm crit}(N_c) = {11 \over 2} N_c + \Big(6 + {4 \over N_c} + {\cal O}(N_c^{-2}) \Big) \epsilon + {\cal O}(\epsilon^2).
\end{align}
In particular, for $SU(2)$ and $SU(3)$ gauge groups, we find $N_f^{\rm crit}(2) = 11 + 8 \epsilon$ and $N_f^{\rm crit}(3) = 33/2 + 7 \epsilon$ to leading order in the $\epsilon$-expansion.
For large $N_c$ (with fixed ratio $N_f/N_c \sim 1$), the estimate in Eq. (\ref{criticalN}) coincides with that obtained from examination of $\beta_g$. However, we see that the large-$N_f$ fixed point is destabilized earlier than might a priori be expected at finite $N_c$.

Previous work has used the $1/N_f$ expansion to study the stability of QCD3.
A solution to the Schwinger-Dyson equations suggests that the theory is driven into a phase in which the fermions acquire a mass at $N_f = {128 \over 3 \pi^2} {N_c^2 - 1 \over N_c}$ \cite{AppelquistNash1990}.
For QCD3 with gauge group SU(2), 
a theory which appears in the study of algebraic spin liquids and theories for high-temperature superconductivity \cite{LeeNagaosaWen2006}, \cite{XuRG4fermi2008} estimates that a particular linear combination of four-fermion operators becomes relevant when $N_f < 6$.
While the $1/N_f$ expansion directly accesses three dimensions, the $\epsilon$-expansion provides a complementary estimate valid for $N_f \sim {\cal O}(1)$.

Our work is inspired by recent studies of three-dimensional quantum electrodynamics QED3 \footnote{We make no distinction between compact $U(1)$ or non-compact $\mathbb{R}$ versions of QED3 here.} 
\cite{FischerAlkoferDahmMaris2004, KavehHerbut2005, StrouthosKogut2009, BraunGiesJanssenRoscher2014, ChesterPufuanomalousdims2016}, in particular the studies via the $\epsilon$-expansion \cite{DiPietroKomargodskiShamirStamou,GiombiKlebanovTarnopolsky}.
(QCD3 may be viewed as a particular ultraviolet completion of compact QED3.)
In these works, it is found that the analogous large-$N_f$ QED3 fixed point \cite{Pisarski1984, AppelquistNashWijewardhanaQED3} becomes unstable for some $1 \leq N_f < 10$.
It is important to understand both in QED3 and QCD3 whether dangerously irrelevant operators drive the theory into a massive phase or to a new massless fixed point \cite{KarthikNarayanan2016} (see \cite{RoscherTorresStrack} for work in this direction in a closely related theory).
We also mention progress on utilizing the conformal bootstrap \cite{DSDTASI2016} to study QED3 \cite{ChesterPufubootstrap2016} and QCD4 \cite{Nakayamabootstrapqcd2016} and leave possible studies of QCD3 to future work.

To complement our study of QCD3 via the $\epsilon$-expansion, we use F-theorem \cite{Jafferis2012, JafferisKlebanovPufuSafdi2011, klebanovpufusafdi2011} or entanglement monotonicity \cite{MyersSinha2011, CasiniHuertaMyers2011, KlebanovPufuSachdevSafdi2012, CasiniHuerta2012, Grover2014} considerations to constrain the possible IR dynamics.
Following \cite{Grover2014}, we show in Sec. \ref{ftheoremsection} that QCD3 admits the flow to a particular massive phase (described in Sec. \ref{qcd3outline}) when there is a solution $N_f^{\rm F}(N_c) \equiv N_f^{\rm F}$ to the equation,   
\begin{align}
\label{FtheoremN}
N^{\rm F}_f N_c F_{\rm dirac} + {N_c^2 - 1 \over 2} \log\Big({\pi N^{\rm F}_f \over 4} \Big) - {N_c^2 - N_c \over 2} \log(2 \pi) - \log(G_2(N_c+1)) = 2 (N^{\rm F}_f)^2 F_{\rm boson},
\end{align}
for fixed $N_c$ where $G_2(z)$ is the Barnes function satisfying $G_2(N_c+1) = 2! 3! \cdots (N_c - 2)(N_c-1)!$.
The constants $F_{\rm dirac} = {\log(2) \over 2} + {3 \zeta(3) \over 4 \pi^2}$ and $F_{\rm boson} = {\log(2) \over 8} - {3 \zeta(3) \over 16 \pi^2}$ are the 3-sphere free energies of a four-component Dirac fermion and real scalar boson with $\zeta(x)$ being the Zeta function.
The left hand-side of Eq. (\ref{FtheoremN}) is valid to ${\cal O}(1/N_f)$ \cite{KlebanovPufuSachdevSafdi2012}; consequently, any solution $N_f^{\rm F}$ -- signifying the critical number of flavors for which a flow from QCD3 to a massive phase is possible -- should be understood to be an estimate valid within the $1/N_f$ expansion.
For $SU(2)$ and $SU(3)$ gauge groups, we find $N_f^{\rm F}(2) \approx 7$ and $N_f^{\rm F}(3) \approx 12$, in close agreement with previous studies of QCD3 via the $1/N_f$ expansion [\onlinecite{AppelquistNash1990, XuRG4fermi2008}].  

$N_f^{\rm F}$ (and the estimates for the critical number of flavors obtained earlier via the  $1/N_f$ expansion [\onlinecite{AppelquistNash1990, XuRG4fermi2008}]) is roughly half the value of $N_f^{\rm crit}$ that we find using the $\epsilon$-expansion in Eq. (\ref{criticalN}) extrapolated to $\epsilon = 1$.
There is no contradiction here, as the domain of validity of the two expansions need not overlap.
Furthermore, $N_f^{\rm F}$ and $N_f^{\rm crit}$, strictly speaking, have different meanings: $N_f^{\rm F}$ signifies when QCD3 is allowed to flow to a massive phase, while $N_f^{\rm crit}$ denotes the point where a particular four-fermion operator becomes relevant.
It is conceivable that the four-fermion operator that is found to destabilize the large-$N_f$ QCD3 fixed point within the $\epsilon$-expansion instead drives the theory to a non-trivial IR fixed point for some range of $N_f^{\rm F} < N_f < N_f^{\rm crit}$ before the massive phase becomes available for $N_f \leq N_f^{\rm F}$ (other possibilities, e.g., the extension of the massless phase to $N_f = 1$, exist as well). 

The remainder of this note is organized as follows. In Section \ref{qcd3outline}, we frame our study of QCD3 within the $\epsilon$-expansion and summarize our conventions.
In Section \ref{dangerousirrelevants}, we present our calculation of the anomalous dimensions of the four-fermion operators in QCD3 from which we derive an estimate for $N_f^{\rm crit}$. In Section \ref{ftheoremsection}, we discuss the estimate of $N_f^{\rm F}$ obtained from F-theorem considerations. We conclude in Section \ref{conclusion}.
There are two appendices that provide further details used in the calculation of Section \ref{dangerousirrelevants}: Appendix \ref{matrixappendix} contains the matrix of anomalous dimensions for the four-fermion operators that we study; Appendix \ref{nomixingappendix} provides details of the argument that there is no mixing of operators that vanish on-shell into those that do not.

\section{QCD3 Preliminaries}
\label{qcd3outline}


We study QCD in three spacetime dimensions via the $\epsilon$-expansion about four dimensions.
For convenience, we generally refer to QCD in $4 - \epsilon$ dimensions with $\epsilon > 0$ as QCD3. 
We take QCD3 to have gauge group $SU(N_c)$ and $N_f$ four-component Dirac spinors $\Psi_n$ ($n = 1, \ldots, N_f$) in the fundamental representation of $SU(N_c)$.
Our aim is to better understand the IR dynamics of the theory as $N_f$ is varied for fixed $N_c$.

In $4 - \epsilon$ dimensions, the QCD action,
\be
\label{action}
S=\int d^{4-\epsilon}x \left(\bar{\Psi}i(\slashed{\pd}-ig\slashed{A}^at^a)\Psi-\frac{1}{4} F^a_{\mu\nu}F_a^{\mu\nu}\right),
\ee
where $F^a_{\mu \nu}$ is the field strength of the gauge field $A_\mu^a$ and $\{t^a\}$ are the generators of $SU(N_c)$ (the sum over $a$ is understood; the sum over the flavor index $n$ and color indices are suppressed).
As usual $\slashed{\pd} \equiv \partial_\mu \gamma^\mu$ and $\slashed{A^a} \equiv A_\mu^a \gamma^\mu$, and  $\bar{\Psi} \equiv \Psi^\dagger \gamma^0$ for $\gamma$-matrices satisfying $\{\gamma^\mu, \gamma^\nu\} = 2 \eta^{\mu \nu}$ with $\eta^{\mu \nu} \eta_{\mu \nu} = 4 - \epsilon$ (see \cite{PeskinSchroeder} and references therein for a discussion of $\gamma$-matrices in non-integral dimension).

In four dimensions ($\epsilon = 0$), the QCD Lagrangian enjoys the global chiral symmetry $SU(N_f) \times SU(N_f) \times U(1)$ in addition to the discrete spacetime symmetries of charge conjugation, time-reversal, and parity.
In three dimensions ($\epsilon = 1$), the ``chiral symmetry" is enhanced to $SU(2 N_f)$; the parity operation becomes reflection along one spatial coordinate with the other two discrete transformations unchanged.
The enhancement of the global symmetry can be understood by writing the $N_f$ four-component Dirac spinors in terms of $2N_f$ two-component spinors $\Psi_n = \begin{pmatrix} \psi_n & \psi_{n + N_f}\end{pmatrix}^T$.
Given our interest in the three-dimensional theory, we will think of the global symmetry of Eq. (\ref{action}) as $SU(2N_f)$.

For $\epsilon > 0$, the gauge coupling $g$ acquires positive mass dimension (at the classical level) and consequently flows towards strong coupling in the IR.
For $\epsilon = 1$, this flow can be reliably studied via the $1/N_f$ expansion.
The leading order in the $\epsilon$-expansion $\beta_g$-function for the gauge coupling -- given in Eq. (\ref{gaugebeta}) -- indicates a non-trivial perturbative fixed point for sufficiently large $N_f \gg {11 \over 2} N_c$ and $\epsilon > 0$.
In fact, this ``large-$N_f$ fixed point" persists and remains perturbative for $0 < \epsilon \ll 1$ as long as $N_f > {11 \over 2} N_c$.
The large-$N_f$ fixed point is the extrapolation to three dimensions of the free fixed point of the IR free phase of four-dimensional QCD.
(A higher-order study in the $\epsilon$-expansion is required to address the behavior of the theory for $N_f < {11 \over 2} N_c$ and $\epsilon > 0$ where the zero of $\beta_g$ at $g^2_\ast$ is lifted.)
Thus, the $\epsilon$-expansion furnishes an alternative method by which to study QCD3 with $N_f \sim {\cal O}(1)$.

The fate of this large-$N_f$ fixed point as $N_f$ is lowered is the subject of this paper and previous studies \cite{AppelquistNash1990, XuRG4fermi2008}.
One hypothesis is that $SU(2 N_f)$ is spontaneously broken to $SU(N_f) \times SU(N_f) \times U(1)$ as $N_f$ is lowered beyond some critical value (the simplest scenario is one in which the discrete symmetries are preserved).
In (three-dimensional) two-component spinor notation, this symmetry breaking can be achieved by a non-zero vacuum expectation value of $\bar{\psi}_n \psi_n - \bar{\psi}_{n + N_f} \psi_{n + N_f}$. 
The precise dynamics that might give rise to such a symmetry breaking is not currently understood, although estimates based upon the $1/N_f$ expansion (\cite{ChesterPufuanomalousdims2016} and references therein) and the $\epsilon$-expansion \cite{DiPietroKomargodskiShamirStamou} in QED3 indicate that a four-fermion operator can become relevant as $N_f$ is lowered and thereby precipitate the symmetry breaking.
However, such higher-body operators need not result in symmetry breaking; they could instead generate the flow to a new non-trivial fixed point.

In Sec. \ref{dangerousirrelevants}, we extend a previous study \cite{DiPietroKomargodskiShamirStamou} of four-fermion operators in QED3 to QCD3 using the $\epsilon$-expansion.
We thereby determine the critical number of flavors $N_f^{\rm crit}(N_c) \equiv N_f^{\rm crit}$ below which a certain linear combination of four-fermion operators becomes relevant and destabilizes the large-$N_f$ fixed point.
Unfortunately, we are unable to determine the endpoint of the resulting renormalization group (RG) flow; we do not know whether the four-fermion operator leads to the spontaneous breaking of $SU(2N_f)$ or if a new non-trivial IR fixed point is achieved.
We leave a more detailed investigation of this important question for further study.
However, we can use the F-theorem to determine when spontaneous symmetry breaking becomes possible and do so in Sec. \ref{ftheoremsection}.


\section{Dangerously Irrelevant Operators in QCD3}
\label{dangerousirrelevants}

In this section, we calculate the anomalous dimensions of $SU(2N_f)$ symmetry-preserving four-fermion operators.
We begin with a summary of our Feynman rules and then discuss the calculation.

\subsection{Feynman rules}

In the computations outlined in this section, we work in Feynman gauge, which is implemented by adding the standard gauge-fixing term to the QCD3 action Eq. \eqref{action},
\be
\Lag_{\text{gauge fixing}}=-\frac{1}{2\xi}(\pd^\mu A^a_\mu)^2.
\ee
Feynman gauge is defined as fixing $\xi=1$. This results in a gauge boson propagator, 
\be
D_{\mu\nu,ab}(p)=
\begin{gathered}
\includegraphics[width=0.1\textwidth]{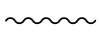}
\end{gathered}
=\frac{-i\eta_{\mu\nu}}{p^2}\delta_{ab},
\ee
where $a$ and $b$ are gauge group indices. Our fermion propagator,
\be
G_{mn, ij}(p)=
\begin{gathered}
\includegraphics[width=0.1\textwidth]{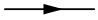}
\end{gathered}
=\frac{i\slashed{p}}{p^2}\delta_{mn} \delta_{ij},
\ee
is obtained directly from the action, where $m, n$ are flavor indices and $i, j$ are color indices. Similarly, the fermion-gauge boson vertex is 
\be
\begin{gathered}
\includegraphics[width=0.1\textwidth]{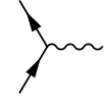}
\end{gathered}
=ig\gamma^\mu t^a\delta_{mn}
\ee
where the flavor indices $m$ ($n$) are attached to the in-coming (out-going) fermion lines.

\subsection{Four-fermion operators}
\label{fourfermionsection}

Following the intuition of \cite{DiPietroKomargodskiShamirStamou,ChesterPufuanomalousdims2016, XuRG4fermi2008}, our interest will be in four-fermion operators. In the three-dimensional theory, one can construct at most four linearly independent four-fermion operators which share the symmetries of the action and, therefore, can mix under the RG. In two-component spinor notation, these are
\bea
\mathcal{O}_V&=&(\bar{\psi}_i\sigma^\mu t_{ij}^a\psi_j)(\bar{\psi}_k\sigma_\mu t_{kl}^a\psi_l) \\
\mathcal{O}_A&=&(\bar{\psi}_i t_{ij}^a\psi_j)(\bar{\psi}_k t_{kl}^a\psi_l)\\
\mathcal{O}_{V'}&=&(\bar{\psi}_i\sigma^\mu t_{ij}^a\psi_l)(\bar{\psi}_k\sigma_\mu t_{kl}^a\psi_j) \\
\mathcal{O}_{A'}&=&(\bar{\psi}_i t_{ij}^a\psi_l)(\bar{\psi}_k t_{kl}^a\psi_j),
\eea
where $i,j,k,l$ are color indices, the Pauli $\sigma$-matrices furnish the (minimal) Clifford representation in three dimensions, and parentheses indicate spinors with contracted flavor indices. One can check that other possible four-fermion operators, such as $(\bar{\psi}_i\sigma^\mu t^a_{ij}T^B\psi_j)^2$, where the $\{T^B\}$ are the generators of the $SU(2N_f)$ flavor group, can be constructed from linear combinations of these four operators.

To translate these operators into the language of four-component spinors in $4-\epsilon$ dimensions, we note that in three dimensions, $\gamma_{[\mu}\gamma_{\nu}\gamma_{\rho]}$ (the square bracket denotes anti-symmetrization over the indices $\mu, \nu, \rho$) is proportional to the identity. 
Thus, inserting the ``identity" into $\mathcal{O}_A$ and $\mathcal{O}_{A'}$ and using the fact that 
$(\gamma^{[\mu}\gamma^\nu\gamma^{\rho]})_{\alpha \beta} (\gamma_{[\mu}\gamma_\nu\gamma_{\rho]})_{\gamma \delta} = (\gamma^\mu\gamma_5)_{\alpha \beta} (\gamma_\mu\gamma_5)_{\gamma \delta}$
in four dimensions, we can write down the four-component spinor analogues of these operators in $4-\epsilon$ dimensions, 
\bea
\label{eq: O_V}
\mathcal{O}_V&=&(\bar{\Psi}_i\gamma^\mu t_{ij}^a\Psi_j)(\bar{\Psi}_k\gamma_\mu t_{kl}^a\Psi_l) \\
\mathcal{O}_A&=&(\bar{\Psi}_i\gamma^\mu\gamma_5 t_{ij}^a\Psi_j)(\bar{\Psi}_k\gamma_\mu\gamma_5 t_{kl}^a\Psi_l) \\
\mathcal{O}_{V'}&=&(\bar{\Psi}_i\gamma^\mu t_{ij}^a\Psi_l)(\bar{\Psi}_k\gamma_\mu t_{kl}^a\Psi_j)\\
\mathcal{O}_{A'}&=&(\bar{\Psi}_i\gamma^\mu\gamma_5 t_{ij}^a\Psi_l)(\bar{\Psi}_k\gamma_\mu\gamma_5 t_{kl}^a\Psi_j). 
\label{eq: O_A'}
\eea
We see immediately that the first two of these operators are the square of the vector and axial currents (thus the subscripts $V$ and $A$). The remaining two operators consist of the two alternate ways of forming color singlets.

While the above operators can generally mix among themselves under the RG, other operators sharing their engineering dimension of $6-2\epsilon$ which are invariant under the same symmetries can mix with them as well. One can construct at most two such operators that are linearly independent. 
We choose them such that they are proportional to the classical equations of motion resulting from Eq. (\ref{action}) and, therefore, vanish on-shell:
\bea
\label{eq: EoM1}
\mathcal{O}_{EoM,1}&=&(\bar{\psi}_i\gamma^\mu t_{ij}^a\psi_j)(\frac{1}{g}[D_\nu,F^{\mu\nu,a}]-\bar{\psi}_i\gamma^\mu t_{ij}^a\psi_j) \\
\mathcal{O}_{EoM,2}&=&\frac{1}{g}[D_\nu,F^{\mu\nu,a}](\frac{1}{g}[D_\nu,F^{\mu\nu,a}]-\bar{\psi}_i\gamma^\mu t_{ij}^a\psi_j).
\label{eq: EoM2}
\eea
Our choice of operators in Eqs. \eqref{eq: EoM1} - \eqref{eq: EoM2} is motivated by the absence of mixing into the operators in Eqs. \eqref{eq: O_V} - \eqref{eq: O_A'} under the RG.
See Appendix \ref{nomixingappendix} for further details on the argument that establishes this result.

\subsection{Anomalous dimensions}

In order to determine whether there exists a linear combination of the operators in Eqs. \eqref{eq: O_V} - \eqref{eq: O_A'} and \eqref{eq: EoM1} - \eqref{eq: EoM2} which become relevant at some $N_f^{\rm crit}(N_c)$, we study the matrix of anomalous dimensions $\gamma$ for these operators to leading order in the $\epsilon$-expansion evaluated at the large-$N_f$ fixed point. Because the operators which vanish on-shell cannot mix under the RG into the operators which do not, we know that $\gamma$ is a block triangular $6\times6$ matrix
\be
\label{eq: full anom dim matrix}
\gamma^{T}=\frac{g_*^2}{16\pi^2}\matleft{ll}\mathcal{A} & 0 \\ \mathcal{B} & \mathcal{C}\matright,
\ee
where we work with the transpose for convenience. The matrix $\mathcal{A}$ corresponds to the mixing of the four-fermion operators in Eqs. \eqref{eq: O_V} - \eqref{eq: O_A'} amongst themselves, and is therefore $4\times 4$. The upper-right block is $0$, as it corresponds to the mixing of the operators in Eqs. \eqref{eq: EoM1} - \eqref{eq: EoM2}, which vanish on-shell, into the four-fermion operators. Finally, the blocks $\mathcal{B}$ and $\mathcal{C}$ respectively correspond to the mixing of the four-fermion operators into the operators which vanish on-shell and the mixing of the operators in Eqs. \eqref{eq: EoM1} - \eqref{eq: EoM2} into themselves. They are nonzero in general.

Because $\gamma$ is block triangular, it will suffice to focus only the block $\mathcal{A}$, as its eigenvalues will also be eigenvalues of $\gamma$ as a whole. This means that we can neglect the mixing of four-fermion operators into the operators which vanish on-shell, a fact that greatly simplifies our computation.



\begin{figure}
\includegraphics[width=0.2\textwidth]{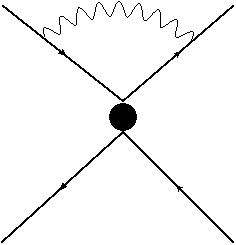}
\includegraphics[width=0.21\textwidth]{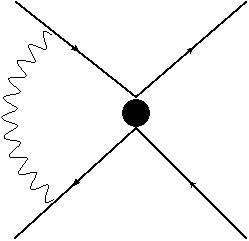}
\includegraphics[width=0.22\textwidth]{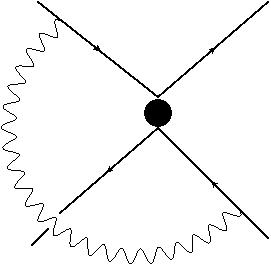} \\
\includegraphics[width=0.25\textwidth]{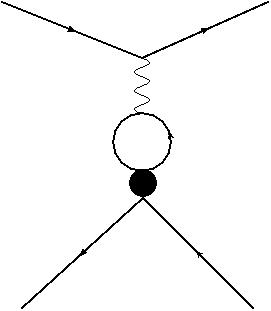}
\includegraphics[width=0.25\textwidth]{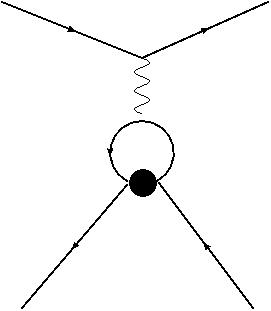}
\caption{The diagrams contributing to the renormalization of the four-fermion operators in Eqs. \eqref{eq: O_V} - \eqref{eq: O_A'} at one-loop. The dot indicates the insertion of a four-fermion operator.}
\label{fig: four-fermi diagrams}
\end{figure}

\begin{figure}
\includegraphics[width=\textwidth]{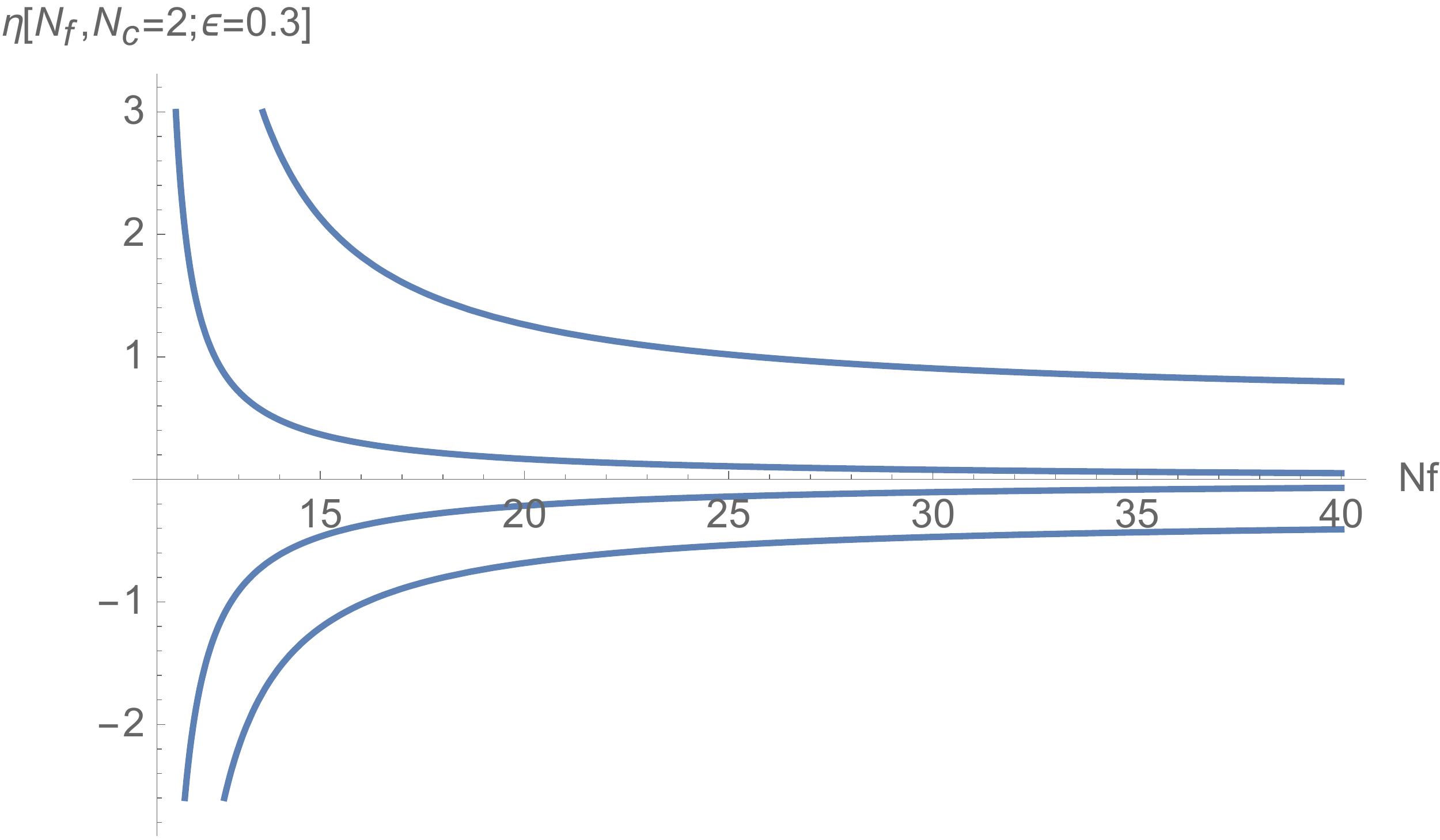}
\caption{The eigenvalues of the block $\frac{g_*^2}{16\pi^2}\mathcal{A}$ of the anomalous dimension matrix corresponding to the mixing of the four-fermion operators in Eqs. \eqref{eq: O_V} - \eqref{eq: O_A'} amongst themselves. Here we take the gauge group to be $SU(2)$ and $\epsilon=0.3$. The large-$N_f$ fixed point of QCD3 becomes unstable when the most negative eigenvalue drops below approximately $-2+0.3=-1.7$.}
\label{fig: SU(2) eigenvalues}
\end{figure}

The block $\frac{g_*^2}{16\pi^2}\mathcal{A}$ of the anomalous dimension matrix can be computed from the diagrams in Fig. \ref{fig: four-fermi diagrams} and is given in Appendix \ref{matrixappendix}. Note that in computing these diagrams we take the external fermion legs in these diagrams to be on-shell. $\frac{g_*^2}{16\pi^2}\mathcal{A}$ has four eigenvalues that correspond to the anomalous dimensions (of the four linear combinations) of four-fermion operators that diagonalize the RG flow.
Two of these eigenvalues are positive and two are negative in the regime we are studying, $N_f>\frac{11}{2}N_c$. In the large-$N_f$ limit, in general one positive and one negative eigenvalue go to zero, while the remaining two asymptote to positive and negative nonzero values.
See Figure \ref{fig: SU(2) eigenvalues} for a plot of these anomalous dimensions for $SU(2)$ gauge group. 

Destabilization of the large-$N_f$ fixed point can only occur when one of the negative anomalous dimensions $\eta(N_f,N_c;\epsilon)$ renders its corresponding operator relevant.
This occurs when
\be
\Delta+\eta(N_f,N_c;\epsilon)<4-\epsilon,
\ee
where $\Delta=6-2\epsilon$ is the engineering dimension of four-fermion operators in $4-\epsilon$ spacetime dimensions. 
Thus, we obtain the condition that an operator is relevant when
\be
\label{eq: dangerously irrelevant}
\eta(N_f,N_c;\epsilon)<-2+\epsilon.
\ee
The large-$N_f$ fixed point becomes unstable when the number of flavors is lowered past the value $N_f^{\rm crit}$ at which this inequality is saturated by at least one eigenvalue $\eta$. Since $\mathcal{A}$ is a $4\times 4$ matrix, obtaining a value of $N_f^{\rm crit}$ analytically is difficult and would likely be unenlightening. Therefore, we estimate $N_f^{\rm crit}$ by diagonalizing $\mathcal{A}$ and solving $\eta(N_f^{\rm crit},N_c;\epsilon)=-2+\epsilon$ for $N_f^{\rm crit}$ given many fixed values of $N_c$ and $\epsilon \ll 1$ \footnote{Specifically, we sample values of $\epsilon$ from $\epsilon=0.01$ to $\epsilon=0.2$ in steps of size $0.005$. We sample each integer $N_c$ from $N_c=2$ to $N_c=30$. Note that the numerical coefficients in Eq. \eqref{result} are rounded to the nearest integer.}. We then fit the result to a linear function of $\epsilon$ for each value of $N_c$, a very good approximation for $\epsilon \ll 1$. We then fit these results as functions of $N_c$. 
From this we find that the first eigenvalue of ${\cal A}$ saturates the bound in Eq. \eqref{eq: dangerously irrelevant} when $N_f$ is lowered to
\be
\label{result}
N_f^{\rm crit}\approx\frac{11}{2}N_c+\left(6+\frac{4}{N_c}+\mathcal{O}(N_c^{-2})\right)\epsilon+\mathcal{O}(\epsilon^2).
\ee
This is an approximate result for the number of flavors below which the fixed point becomes unstable. 
Eq. \eqref{result} implies, in particular, that as $N_c$ becomes larger, $N_f^{\rm crit}$ nears the boundary of the QCD4 conformal window, $N_f=\frac{11}{2}N_c$; this point marks the boundary above which the one-loop beta function of four-dimensional QCD admits an IR free phase (as a function of $N_f$ for fixed $N_c$).

It is an open question as to the fate of the three-dimensional theory for $N_f \leq N_f^{\rm crit}$.
It is possible that the theory flows to the three-dimensional version of the Banks-Zaks fixed point \cite{BanksZaks1982} -- which appears at two-loops in four-dimensional QCD -- before the theory becomes asymptotically free (with respect to the first quantum correction to $\beta_g$) and chiral symmetry is presumably broken. 
To gain a better understanding of when symmetry breaking can occur for $N_f<N_f^{\rm crit}$, in the next section we present an upper bound on the number of flavors below which chiral symmetry can be maximally broken using the F-theorem.

\section{F-theorem and Entanglement Monotonicity}
\label{ftheoremsection}

In the previous section, we determined the critical value of $N^{\rm crit}_f(N_c$) below which a potentially-destabilizing four-fermion interaction became relevant using the $\epsilon$-expansion.
We now consider a complementary perspective from which to assess the fate of QCD3 as the number of flavors is lowered.
We use the F-theorem \cite{Jafferis2012, JafferisKlebanovPufuSafdi2011, klebanovpufusafdi2011} or entanglement monotonicity \cite{MyersSinha2011, CasiniHuertaMyers2011, KlebanovPufuSachdevSafdi2012, CasiniHuerta2012, Grover2014} -- valid for conformal field theories in three spacetime dimensions -- to determine the maximal number of flavors $N_f^{\rm F}(N_c)$ below which the large-$N_f$ QCD3 fixed point may flow to a particular phase in which the fermions acquire a mass, following the idea presented in [\onlinecite{Grover2014}].
In short, for $N_f > N_f^{\rm F}$, the large-$N_f$ fixed point is stable to symmetry breaking; for $N_f < N_f^{\rm F}$, the instability becomes available.

Our analysis assumes a pattern of symmetry breaking in which the possible dynamically-generated fermion masses preserve the $SU(N_f) \times SU(N_f) \times U(1) \subset SU(2 N_f)$ subgroup of the global flavor symmetry, consistent with [\onlinecite{VafaWitten1984}].
(Other types of symmetry breaking are possible, however, they will not be considered here.)
Goldstone's theorem says that the spontaneous symmetry breaking $SU(2N_f) \mapsto SU(N_f) \times SU(N_f) \times U(1)$ results in $2N_f^2$ (real) massless scalars.
Asymptotic freedom then implies that the Goldstone bosons saturate the low-energy field content.

The F-theorem admits RG flow from QCD3 to the (massive) Goldstone phase when
\begin{align}
\label{inequality}
F_{{\rm QCD}3} > F_{\rm Goldstone},
\end{align}
where $F_{{\rm QCD}3}$ and $F_{\rm Goldstone}$ denote the free energies of the two theories on the 3-sphere.
The values of these 3-sphere free energies can be found in [\onlinecite{KlebanovPufuSachdevSafdi2012}]:
\begin{align}
\label{Fs}
F_{{\rm QCD}3} & = N_c N_f \Big( {\log(2) \over 2} + {3 \zeta(3) \over 4 \pi^2}\Big) + {N_c^2 - 1 \over 2} \log\Big({\pi N_f \over 4} \Big) - {N_c (N_c - 1) \over 2} \log(2\pi) \cr
& - \log(G_2(N_c+1)) +\ldots,\cr
F_{\rm Goldstone} & = 2 N_f^2\Big( {\log(2) \over 8} - {3 \zeta(3) \over 16 \pi^2}\Big),
\end{align}
where the $\ldots$ represent additional contributions to $F_{{\rm QCD}3}$ that are expected to begin at ${\cal O}(1/N_f)$. 
$G_2(z)$ is the Barnes function satisfying $G_2(N_c+1) = 2! 3! \cdots (N_c - 2)(N_c-1)!$.

\begin{figure}
\includegraphics[width=\textwidth]{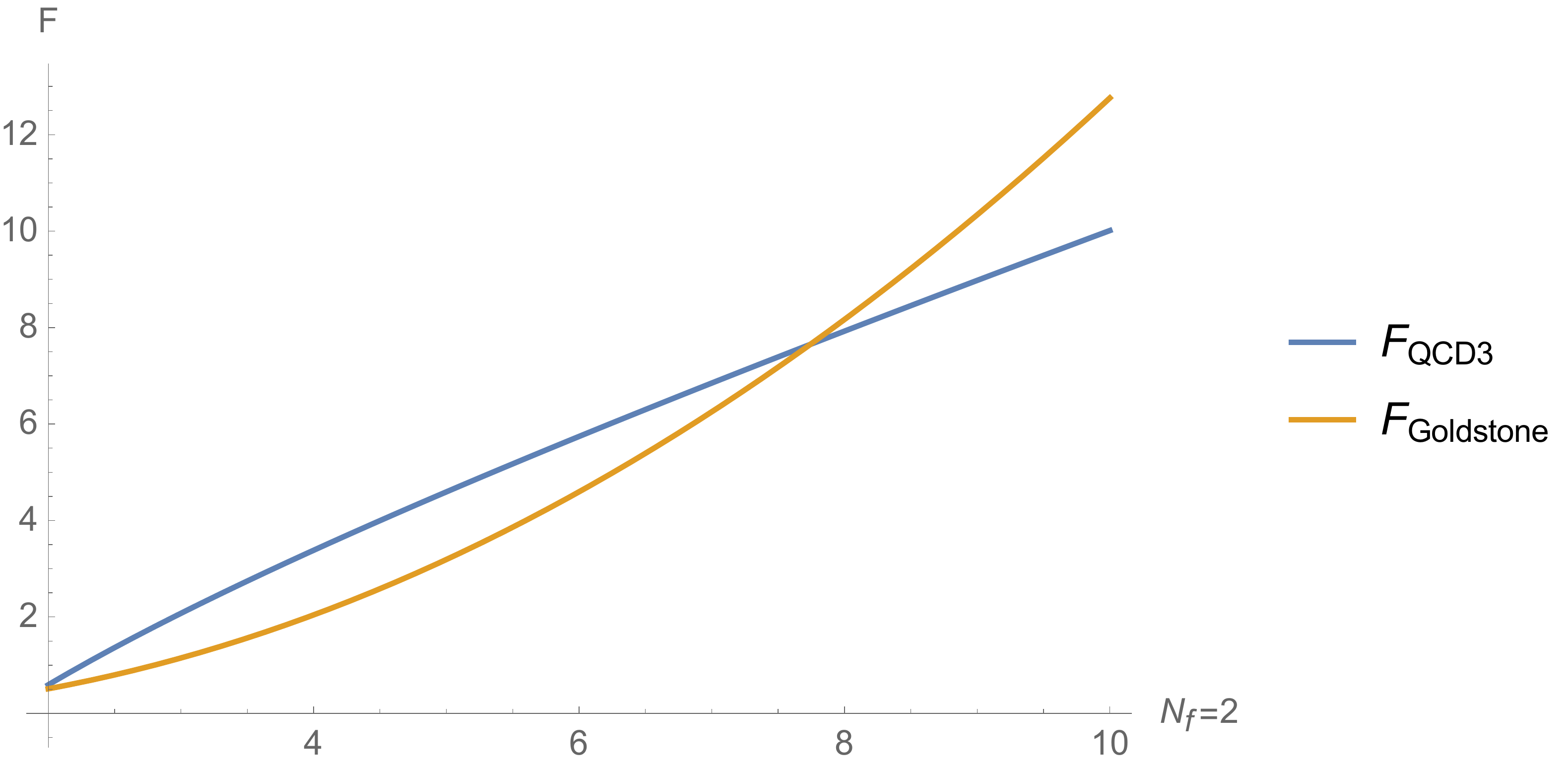}
\caption{The free energies $F_{\rm QCD3}$ (blue) and $F_{\rm Goldstone}$ (yellow) for $SU(2)$ QCD3 as functions of the number of fermion flavors $N_f$. Notice that they cross at $N_f^{\rm F}\approx 7.7$. Note that the origin is placed at $N_f=2$.}
\label{fig: SU(2) free energies}
\end{figure}

A plot of $F_{\rm QCD3}$ for the gauge group $SU(2)$ and $F_{\rm Goldstone}$ is given in Figure \ref{fig: SU(2) free energies}. 
We see that $F_{\rm Goldstone} > F_{\rm QCD3}$ for $N_f \geq 8$, while the Goldstone phase becomes available for smaller $N_f$.

We have not found it possible to analytically solve for the point at which the inequality in Eq. (\ref{inequality}) is saturated.
However, we can estimate $N_f^{\rm F}$ as a function of $N_c$ by numerically minimizing $|F_{\rm QCD}-F_{\rm Goldstone}|$ for many values of $N_c$ and fitting the result \footnote{Here we sample integer values of $N_c$ from $N_c=2$ to 60.}. This yields the estimate
\be
\label{eq: numerical N_f^F}
N_f^{\rm F}\approx4.24N_c-0.35.
\ee
This estimate is very good for large values of $N_c$, for which it gives $\frac{|F_{\rm QCD3}-F_{\rm Goldstone}|}{F_{\rm QCD}}\approx 0.1\%$. It is somewhat worse for smaller values of $N_c$. For example, for the gauge groups $SU(2)$ and $SU(3)$, $\frac{|F_{\rm QCD3}-F_{\rm Goldstone}|}{F_{\rm QCD3}}\approx 5\%$ and $2\%$ respectively. For these gauge groups, we find $N_f^{\rm F}(2)\approx7.7$ and $N_f^{\rm F}(3)\approx12.1$ without performing any fitting.

Because the $1/N_f$ expansion was required to obtain $F_{\rm QCD3}$ above, it is useful to consider alternative means of estimating the large-$N_f$ QCD3 3-sphere free energy.
(It would be interesting to generalize to QCD3 the technique used in \cite{GiombiKlebanovTarnopolsky} to compute $F_{\rm QED3}$ within an $\epsilon$-expansion about four dimensions in order to provide a more direct comparison to $N_f^{\rm crit}$ computed in the previous section.) 
One option is ${\cal N}=2$ supersymmetric QCD3 (SQCD3) whose 3-sphere free energy can be found exactly using localization techniques [\onlinecite{KapustinWillettYaakov2010}]. 
The SQCD3 free energy provides an upper bound on the large-$N_f$ QCD3 free energy since the former flows to the latter under suitable deformation.
Unfortunately, we do not find a lower value of $N_f^{\rm F}(N_c)$. 
For example, for $SU(2)$ and $SU(3)$ gauge groups, we find $N_f^{\rm F}(2) \approx 13$ and $N_f^{\rm F}(3) \approx 18$ using ${\cal N} = 2$ SQCD3.

We remark that it is not helpful to use the 3-sphere free energy of the decoupled UV limit of QCD3.
The reason is that $N_c^2 - 1$ (abelian) gauge fields do not define a conformally invariant theory in three dimensions.
Their free energy scales logarithmically with the radius of the 3-sphere and, therefore, diverges at long wavelengths \cite{KlebanovPufuSachdevSafdi2012}.

\section{Discussion}
\label{conclusion}

In this paper, we utilized the $\epsilon$-expansion about four spacetime dimensions to estimate an upper bound on the number of flavors below which the large-$N_f$ QCD3 is destabilized. 
This was done by finding the number of flavors, Eq. (\ref{result}), at which a certain linear combination of four-fermion operators becomes relevant. 
In addition, we used the F-theorem or entanglement monotonicity to estimate in Eq. (\ref{eq: numerical N_f^F}) when the large-$N_f$ fixed point admits the spontaneous symmetry breaking $SU(2N_f) \mapsto SU(N_f) \times SU(N_f) \times U(1)$.
 
Our computations in Sec. \ref{dangerousirrelevants} were done entirely at the one-loop level. 
It would be of great interest to study this problem out to two-loops in the future.
This might enable one to develop an understanding of the fate of the Banks-Zaks fixed point \cite{BanksZaks1982} in four-dimensional QCD when it is continued to three dimensions. 

The possible applications of QCD3 range from the physics of high-temperature (four-dimensional) QCD \cite{AppelquistPisarski1981} to theories of high-temperature superconductivity \cite{LeeNagaosaWen2006} as well as to magnetic systems \cite{XuRG4fermi2008} and parton descriptions of the quantum Hall effect \cite{Wennonabelian1991}.
We hope that our work may be helpful to such applications.

\acknowledgments

We thank Srinivas Raghu for greatly contributing to the development of this work and for many useful discussions. 
We also thank Lorenzo Di Pietro, Ethan Dyer, Michael Peskin, and Cenke Xu for helpful discussions and correspondence.
M.M. is grateful for the generous hospitality of the Kavli Institute for Theoretical Physics in Santa Barbara where this work was completed.
This research is supported in part by the National Science Foundation Graduate Research Fellowship Program under Grant No. DGE-1144245
(H.G.), the John Templeton Foundation (M.M.), and the National Science Foundation under Grant No. NSF PHY11-25915 (M.M.)

\appendix

\section{Anomalous Dimension Matrix}
\label{matrixappendix}

Computing the diagrams in Figure \ref{fig: four-fermi diagrams}, we obtain the block of the anomalous dimension matrix in Eq. \eqref{eq: full anom dim matrix} corresponding to the mixing of the four-fermion operators in Eqs. \eqref{eq: O_V} - \eqref{eq: O_A'} with themselves
\be
\frac{g_*^2}{16\pi^2}\mathcal{A}_{IJ},
\ee
where $I,J=V,A,V',A'$. The entries of this matrix are
\bea
\mathcal{A}_{VV}&=&\frac{16}{3}T_2(F)N_f+\frac{38}{3}C_2(F)-\frac{10}{3}C_2(G)-5 \\
\mathcal{A}_{VA}&=&6(C_2(F)+\frac{1}{2})\\
\mathcal{A}_{VV'}&=&5(1+\frac{1}{N_c})\\
\mathcal{A}_{VA'}&=&3(1-\frac{1}{N_c})\\
\mathcal{A}_{AV}&=&6(C_2(F)+\frac{1}{2})+\frac{8}{3}(C_2(F)-\frac{1}{2}C_2(G))\\
\mathcal{A}_{AA}&=&10(C_2(F)-\frac{1}{2})-2C_2(G) \\
\mathcal{A}_{AV'}&=&\mathcal{A}_{VA'}\\
\mathcal{A}_{AA'}&=&5(1+\frac{1}{N_c}) \\
\mathcal{A}_{V'V}&=&5(1+\frac{1}{N_c})+\frac{16}{3}(C_2(F)-\frac{1}{2}C_2(G))N_f+\frac{8}{3}T_2(F)\\
\mathcal{A}_{V'A}&=&\mathcal{A}_{VA'}\\
\mathcal{A}_{V'V'}&=&\frac{16}{3}(C_2(F)-\frac{1}{2}C_2(G))N_f+10C_2(F)-2C_2(G)+\frac{8}{3}T_2(F)-5 \\
\mathcal{A}_{V'A'}&=&\mathcal{A}_{VA}\\
\mathcal{A}_{A'V}&=&3(1-\frac{1}{N_c})+\frac{8}{3}T_2(F)\\
\mathcal{A}_{A'A}&=&\mathcal{A}_{AA'}\\
\mathcal{A}_{A'V'}&=&\mathcal{A}_{VA}\\
\mathcal{A}_{A'A'}&=&10(C_2(F)-\frac{1}{2})-2C_2(G)
\eea
where $T_2(F)=\frac{1}{2}$, $C_2(F)=\frac{N_c^2-1}{2N_c}$, and $C_2(G)=N_c$ ($F$ and $G$ are the fundamental and adjoint representations of $SU(N_c)$ respectively). 

To give the reader an idea of how these matrix elements are computed, we describe a sample calculation. Consider the operator $\lambda_V\mathcal{O}_V$, where $\mathcal{O}_V$ is given in Eq. \eqref{eq: O_V}. The Feynman rule for an insertion of this operator is
\be
\begin{gathered}
\includegraphics[width=0.25\textwidth]{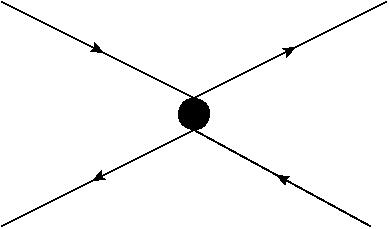}
\end{gathered}
=i\lambda_V\gamma^\mu_{\alpha\beta}t^a_{ij}\gamma_{\mu,\gamma\delta}t^a_{kl}
\ee
where $\alpha,\beta$ ($i,j$) and $\gamma,\delta$ ($k,l$) are the spinor (color) indices associated with the fermion line above and below the dot respectively. As an example of a one-loop insertion of this vertex, consider the second diagram in Figure \ref{fig: four-fermi diagrams}. This diagram is proportional to 
\be
(\gamma^\mu\gamma^\nu\gamma^\rho)_{\alpha\beta}(\gamma_\rho\gamma_\nu\gamma_\mu)_{\gamma\delta}(t^at^b)_{ij}(t^bt^a)_{kl}.
\ee
where we're suppressing the logarithmic divergence. We'll first consider the products of gamma matrices. Using the identity
\be
\gamma^\mu\gamma^\nu\gamma^\rho=\eta^{\nu\rho}\gamma^\mu-\eta^{\mu\rho}\gamma^\nu+\eta^{\mu\nu}\gamma^\rho-i\epsilon^{\mu\nu\rho\sigma}\gamma_\sigma\gamma_5,
\ee
we obtain (using the shorthand $(\gamma^\mu\gamma^\nu\gamma^\rho)_{\alpha\beta}(\gamma_\rho\gamma_\nu\gamma_\mu)_{\gamma\delta}\equiv\gamma^\mu\gamma^\nu\gamma^\rho\otimes\gamma_\rho\gamma_\nu\gamma_\mu$)
\bea
\gamma^\mu\gamma^\nu\gamma^\rho\otimes\gamma_\rho\gamma_\nu\gamma_\mu&=&\gamma^\mu\otimes\gamma^\nu\gamma_\nu\gamma_\mu-\gamma^\nu\otimes\gamma^\mu\gamma_\nu\gamma_\mu+\gamma^\rho\otimes\gamma_\rho\gamma^\mu\gamma_\mu\nonumber\\
&&-i\epsilon^{\mu\nu\rho\sigma}\gamma_\sigma\gamma_5\otimes\gamma_\rho\gamma_\nu\gamma_\mu \\
&=&(2D-(2-D))\gamma^\mu\otimes\gamma_\mu+(i)^2\epsilon^{\mu\nu\rho\sigma}\epsilon_{\rho\nu\mu\delta}\gamma_\sigma\gamma_5\otimes\gamma^\delta\gamma_5 \\
&=&(3D-2)\gamma^\mu\otimes\gamma_\mu+(D-1)!\gamma^\mu\gamma_5\otimes\gamma_\mu\gamma_5,
\eea
where in the second and third lines we used
\bea
\gamma^\mu\gamma_\mu&=&D\\
\gamma^\mu\gamma^\nu\gamma_\mu&=&(2-D)\gamma^\nu \\
\epsilon^{\mu\nu\rho\sigma}\epsilon_{\rho\nu\mu\delta}&=&-(D-1)!\delta^{\sigma}_{\text{ }\delta},
\eea
where $D=4-\epsilon$. We now move on to the product of the gauge group generators. Here we will use the commutator
\be
[t^a,t^b]=if^{abc}t^c
\ee
and the identities
\bea
t^a_{ij}t^a_{kl}&=&\frac{1}{2}(\delta_{il}\delta_{kj}-\frac{1}{N_c}\delta_{ij}\delta_{kl})\\
if^{abc}t^bt^c&=&-\frac{1}{2}C_2(G)t^a.
\eea
We therefore have
\bea
(t^at^b)_{ij}(t^bt^a)_{kl}&=&if^{abc}t^c_{ij}(t^bt^a)_{kl}+(t^bt^a)_{ij}(t^bt^a)_{kl}\\
&=&\frac{1}{2}C_2(G)t^a_{ij} t^a_{kl}+\frac{1}{2}t^a_{mj}t^a_{nl}(\delta_{in}\delta_{km}-\frac{1}{N_c}\delta_{im}\delta_{kn})\\
&=&\left[\frac{1}{2}C_2(G)-\frac{1}{2N_c}\right]t^a_{ij}t^a_{kl}+\frac{1}{2}t^a_{il}t^a_{kj}\\
&=&C_2(F)t^a_{ij}t^a_{kl}+\frac{1}{2}t^a_{il}t^a_{kj},
\eea
where the first term results in mixing into $\mathcal{O}_V$ and $\mathcal{O}_A$, and the second term results in mixing into $\mathcal{O}_{V'}$ and $\mathcal{O}_{A'}$. The remaining diagrams can be computed analogously.

\section{RG Mixing with Redundant Operators}
\label{nomixingappendix}

In computing eigenvalues of the anomalous dimension matrix $\gamma$ in Section III, it was of great use to select a basis of operators $\{\mathcal{O}_I\}$ such that $\gamma$ is block-triangular. This was done by selecting two operators which vanish upon use of the classical equations of motion (the contributions of higher-dimension operators are assumed irrelevant). Such operators are called redundant \cite{WeinbergQFTvol1}. We argue in this appendix that redundant operators in general do not mix into operators which are not redundant under the RG. Much of the argument in this section has overlap with that in \cite{Arzt1995, EinhornWudka2001}. 

A redundant operator is defined as an operator for which infinitesimal variations in its coupling can be eliminated from the action by way of a redefinition of the fields $\{\Psi_i\}$ in the theory \cite{WeinbergQFTvol1}. Such an infinitesimal field redefinition of a field $\Psi_i$ takes the form
\be
\Psi_i\mapsto\Psi_i+\epsilon F(\Psi_j,\pd\Psi_j,...)
\ee
where $F$ is some continuous function of the fields in the theory and their derivatives. The change in the action under this variation is therefore
\be
\delta S=\epsilon\frac{\delta S}{\delta\Psi_i}F(\Psi_j,\pd\Psi_j,...).
\ee
Thus, an operator $\mathcal{O}$ with coupling $\lambda$ is redundant if under $\lambda\mapsto\lambda+\delta\lambda$
\be
\label{eq: redundant operator def}
\frac{\delta S}{\delta\lambda}=-\sum_i\frac{\delta S}{\delta\Phi_i}F_i(\Psi_j,\pd\Psi_j,...)
\ee
for some subset $\{\Phi_i\}$ of the fields. Thus, redundant operators are operators which vanish on-shell (i.e. $\frac{\delta S}{\delta \Psi}=0$). 

As an aside, we can generalize the concept of a redundant operator to that of a redundant parameter. A redundant parameter $\Omega$ in a theory (not necessarily just the coupling constant associated with a single operator) is redundant if $\frac{\delta S}{\delta\Omega}$ takes the form of Eq. \eqref{eq: redundant operator def}, meaning that variations in $\Omega$ can be canceled by a field redefinition. A redundant operator is therefore an operator with a coupling constant which is a redundant parameter. Redundant parameters in general cannot affect observables like $S$-matrix elements, masses, charges, and anomalous dimensions at a RG fixed point. Redundant parameters can, however, appear in RG-dependent quantities like beta functions; a procedure for their removal has been achieved in \cite{Arzt1995, EinhornWudka2001}. 

The special case of a redundant operator is particularly well behaved since, to satisfy Eq. \eqref{eq: redundant operator def}, this operator must vanish on the equations of motion, so an infinitesimal field redefinition can always remove it from the bare action (up to the Jacobian of the redefinition and a shift in the source of the field being redefined, both of which we will discuss below) provided that it is irrelevant. Thus, such operators cannot renormalize operators which are not redundant (note that the converse need not be true). Below, we will give a rough argument for this for the case of interest, $SU(N_c)$ QCD in $4-\epsilon$ dimensions. 

We will be interested in the mixing of dimension-6 (under four-dimensional power counting) operators under the RG which are invariant under the symmetries of the action in Eq. \eqref{action}. As in Sec. \ref{fourfermionsection}, when choosing a basis of these operators, we choose two redundant operators, one of which is (suppressing color indices)
\be
\mathcal{O}_{EoM,1}=\mathcal{J}_V^{\mu}(\frac{1}{g}[D^\nu,F_{\mu\nu}]-\mathcal{J}_{V,\mu}).
\ee
This is the operator in Eq. \eqref{eq: EoM1} written in terms of the vector current $\mathcal{J}_V^{\mu,a}=\bar{\Psi}\gamma^\mu t^a\Psi$ with gauge group indices suppressed. 
The second operator ${\cal O}_{EoM,2}$ may be handled similarly.
The term in the Lagrangian associated with this operator can be written as
\be
\Lag\supset\lambda_{EoM,1}\mathcal{J}_V^\mu\frac{\delta S}{\delta A^\mu}.
\ee
It is certainly true that any change in $\lambda_{EoM, 1}$ under the RG can be removed by way of a field redefinition, but we are primarily interested in the other dimension-6 operators that are not redundant, so we must remove $\lambda_{EoM,1}$ from the bare action in order to keep beta functions of other couplings from depending on it. 
This is possible because ${\cal O}_{EoM,1}$ is irrelevant and so its coupling $\lambda_{EoM,1}$ is naturally proportional to two powers of the inverse cutoff $1/\Lambda^2$.
We then perform the infinitesimal field redefinition
\be
A'^\mu= A^\mu- {1 \over \Lambda^2} \mathcal{J}_V^\mu
\ee
which eliminates $\mathcal{O}_{EoM, 1}$ from the bare action and prevents us from having to worry about its effect on the running of non-redundant couplings in the theory. Of course, under the RG, $\mathcal{O}_{EoM,1}$ can be generated, but, again, it can always be removed in this way at each step in the RG procedure. The point is that because it can be eliminated by way of an infinitesimal field redefinition, $\lambda_{EoM,1}$ cannot contribute to the renormalization of other, non-redundant couplings.

Note that the above redefinition of the gauge field will introduce a Jacobian in the path integral which can be neglected (it can be generally interpreted as introducing ghosts which we can for all intents and purposes ignore). There is also an additional term of the form ${1 \over \Lambda^2} J_\mu\mathcal{J}_V^\mu$, where $J_\mu$ is the source of $A^\mu$, that appears when we take into account source terms, but this term does not affect the four-fermion correlation functions we are primarily interested in. Further discussion can be found in \cite{Arzt1995}.

\bibliography{qcd3stability}

\end{document}